\documentclass [12pt]{article}

\usepackage{graphicx,float}
\usepackage{authblk}
\usepackage{amsmath}
\usepackage{multirow}
\usepackage{array}
\usepackage{color,subfigure,amssymb,enumerate}

\definecolor{maroon}{rgb}{0.5, 0, 0}

\newcommand{\ma}[1]{\textcolor{black}{#1}}

\newcommand{\ie}{{\it i.e.}\ }

\newcommand{\etal}{{\it et al.}\ }

\newcommand{\be}{\begin{eqnarray}}
\newcommand{\ee}{\end{eqnarray}}

\newcommand{\bes}{\begin{eqnarray*}}
\newcommand{\ees}{\end{eqnarray*}}
\newcommand{\ds}{\displaystyle}
\newcommand{\vect}[1]{\mbox{\boldmath $#1$}}

\newcommand{\dd} \partial

\newcommand{\eps}{\epsilon}
\newcommand{\non} \nonumber

\newcommand{\helec}{\dot{\cal Q}^{(\rm elyte)}}
\newcommand{\hama}{\dot{\cal Q}_{\rm part.}^{(\rm a)}}
\newcommand{\hamc}{\dot{\cal Q}_{\rm part.}^{(\rm c)}}
\newcommand{\hohma}{\dot{\cal Q}_{\rm ohm}^{(\rm a)}}
\newcommand{\hohmc}{\dot{\cal Q}_{\rm ohm}^{(\rm c)}}
\newcommand{\hopa}{\dot{\cal Q}_{\rm pol.}^{(\rm a)}}
\newcommand{\hopc}{\dot{\cal Q}_{\rm pol.}^{(\rm c)}}
\newcommand{\hrev}{\dot{\cal Q}_{\rm rev.}}
\newcommand{\hirr}{\dot{\cal Q}_{\rm irr.}}
\newcommand{\wdot}{\dot{\cal W}}

\newcommand{\gibbs}{G}

\newcommand{\tn}{t_+^0}

\newcommand{\cd}{{\cal B}}

\newcommand{\vph}{\varphi}

\newcommand{\N}{{\cal N}}

\newcommand{\cam}{c_a^{\rm max}}
\newcommand{\ccm}{c_c^{\rm max}}
\newcommand{\ueqa}{U_{eq,a}}
\newcommand{\ueqc}{U_{eq,c}}

\newcommand{\mub}{\bar{\mu}}
\newcommand{\mubb}{{\mu}}

\newcommand{\avvj}{\langle j \rangle}
\newcommand{\avvqe}{\langle N_{\rm E_{\rm tot}} \rangle}
\newcommand{\avvqn}{\langle N_- \rangle}
\newcommand{\avvqp}{\langle N_+\rangle}
\newcommand{\avvome}{\langle \omega_e\rangle}

\newcommand{\avjn}{\bar{j}_{\rm n}}

\newcommand{\ev}{\epsilon_l}
\newcommand{\bet}{b_{et}}

\newcommand{\G}{\mathcal{G}}

\newcommand{\ecs}{{G}_{\rm s,part}}
\newcommand{\ecsa}{{G}^{(\rm a)}_{\rm s,part}}
\newcommand{\ecsc}{{G}^{(\rm c)}_{\rm s,part}}

\newcommand{\tp}{t_0^+}
\newcommand{\De}{D_{\rm{e}}}

\usepackage[margin=1in]{geometry}

\title{Heat Generation and a Conservation Law for Chemical Energy in Lithium-ion batteries}
\author[1,2]{G. Richardson}
\author[1,2]{I. Korotkin}
\affil[1]{Mathematical Sciences, University of Southampton, University Rd., SO17 1BJ, UK}
\affil[2]{The Faraday Institution, Quad One, Becquerel Avenue, Harwell Campus, Didcot, OX11 0RA, UK}

\begin{document}

\maketitle

\begin{abstract}

Present theories of irreversible energy losses and heat generation within Li-ion cells are unsatisfactory because they are not compatible with energy conservation. This work aims to provide a consistent theoretical treatment of energy transport and losses in such devices. An energy conservation law  is derived  from the Doyle-Fuller-Newman (DFN) model of a Li-ion cell using a rigorous mathematical approach. The resulting law allows irreversible chemical energy losses to be located to seven different regions of the cell, namely: (i)  the electrolyte, (ii) the anode particles, (iii) the cathode particles, (iv)  the solid parts of the anode (ohmic losses), (v)  the solid parts of the cathode (ohmic losses), (vi)  the surfaces of the anode particles (polarisation losses), and (vii)  the surfaces of the cathode particles (polarisation losses). Numerical solutions to the DFN model are used to validate the conservation law in the cases of a drive cycle and constant current discharges, and to compare the energy losses occurring in different locations. It is indicated how cell design can be improved, for a specified set of operating conditions, by comparing the magnitude of energy losses  in the different regions of the cell.\\
{\bf Keywords: Li-ion battery, Energy conservation Law, Newman model, Heat production}


\end{abstract}

\subsection*{Highlights}
\begin{itemize}
\item We derive an energy conservation law for the Doyle-Fuller-Newman model of a Li-ion cell.
\item We obtain heat production rates for a Li-ion cell that are consistent with energy conservation.
\item We indicate how precise knowledge of energy dissipation within the cell can be used as a design tool.
\end{itemize}


\section{Introduction}


The drive  to eliminate carbon based fuels from transportation systems, and the resulting legislation to phase out the internal combustion engine across large parts of the world before
2040, has led to rapid growth interest in lithium ion battery (LIB) technology. Currently LIB technology is used in most portable consumer electronics, and is increasingly being used in home energy storage units, but it is its use in electric vehicles (EVs) that is set to see the biggest growth in its market, which is set to increase from 45 GWh/year (in 2015) to around 390 GWh/year in 2030 \cite{zubi18}. The prime reason for its dominance of the  automotive industry is  its unrivalled high power and energy densities. It also has the advantages of discharging slowly when not in use, little or no need for maintenance and the ability to undergo a large number of charge/discharge cycles without significant degradation. 

The Doyle Fuller Newman (DFN) model \cite{arora00,doyle93,fuller94b,fuller94,NewmanBook} has proved itself to be an extremely useful, and versatile, tool for understanding lithium-ion battery performance. Recent works have shown that, providing that lithium ion transport within the electrode particles is modelled by an appropriately calibrated nonlinear diffusion equation, the model is capable of accurately predicting battery performance \cite{ecker15}, even when subjected to highly non-uniform drive cycles \cite{alana}.
While these predictions of the DFN model provide an accurate relation between the cell voltage and the current draw they have not, to date, been used to provide a consistent picture of the irreversible energy losses occurring within the cell.  While there are many works that use DFN to estimate irreversible energy loss and heating within lithium-ion cells none of them uses a theory of energy dissipation that is consistent with the DFN  model. In this context we note the following works that are  based on DFN simulation \cite{campbell19,du17,erhard15,fang10,gomadam02,hunt20,kumaresan07,lai15,munoz20,nie21,northrop15,srinivasan02,sturm19,torchio16,wang02} and \cite{baba14}, which is based on a single particle model. All of these works predict energy losses without accounting for the enthalpy of mixing (or heat of mixing) in the electrode particles and only partially account for the irreversible energy loss in the electrolyte, again neglecting the enthalpy of mixing. This method of estimating the energy dissipation is based on  thermodynamic treatments by Rao and Newman \cite{rao97} and Gu and Wang \cite{gu00}, which are ultimately based on the work of Bernardi \cite{bernardi85}. We note also the works Tranter \etal \cite{tranter20} and Farag \etal \cite{farag17}, which are both based on the DFN model, but use alternative methods for estimating heat production, neither of which are consistent with overall energy conservation within the DFN model, although Farag \etal do approximate the heat of mixing within the electrode particles, noting that it is often significant. 
Finally, we remark that Latz and Zausch \cite{latz11,latz15} have used a thermodynamic method to estimate energy dissipation in a lithium ion cell and, as we shall show, obtain expressions for irreversible energy losses within the device that are consistent with the energy conservation law that we derive here from the DFN model. However, as far as we are aware, their work has never been applied to the DFN model. The relative lack of attention that their estimate of energy dissipation has received can be attributed to (I) the large number of other works in the literature that use thermodynamic arguments to arrive at incomplete estimates of the irreversible energy losses and (II) the fact that there is no independent theoretical procedure for determining which of these thermodynamic approaches are correct. The current work aims to address the second of these points and thereby fill a significant void in the literature.

Given the widespread use, and overall utility, of the DFN model of Li-ion battery behaviour it would be a major step forward to unequivocally establish the form of the irreversible energy dissipation law that is consistent with this model. 
This will not only allow accurate modelling of heat generation in composite cells (such as cylindrical and pouch cells), in which inadequate cooling can lead to significant temperature heterogeneities, but could also be used as a design tool in order to identify the components of the cell in which irreversible losses are most significant under the cell's characteristic operational conditions. To date, a unifying theory of energy transport and dissipation in the DFN model remains to be established and it is precisely this omission that we aim to address here. In order to accomplish this we restrict our attention to a single cell, which given its small width, we consider to have spatial uniform temperature $T=T(t)$.  Rather than adopt a thermodynamic approach we directly derive an energy conservation law from the DFN model. This has the advantage that it avoids the pitfalls and intricacies of non-equilibrium thermodynamics, which, in this application, have led to a number of incorrect (or at best approximate) results.  In fact we rigorously prove that the conservation law is a consequence of the DFN model, and hence demonstrate unequivocally that the irreversible energy loss terms derived here are the only ones consistent with the DFN model. This result is validated against numerical solutions to the DFN model, both for constant current discharge and drive cycles. These solutions are computed using the fast, second-order accurate, DFN solver DandeLiion \cite{dande} and are, in turn, used to evaluate each term in the energy conservation equation.

The paper is set out as follows. In \S2 we recap the Doyle-Fuller-Newman model. In \S3 we summarise the energy conservation law in a concise form, which can be easily applied to the DFN model, and then in \S3.2 we validate this law against full simulations of the model, both for constant current discharge and for a drive cycle. The derivation of the energy conservation law from the DFN model is presented in \S4 before finally, in \S5, we draw our conclusions.

\section{The Doyle-Fuller-Newman model}

\begin{figure}[ht]
\begin{center}
\scalebox{0.6}{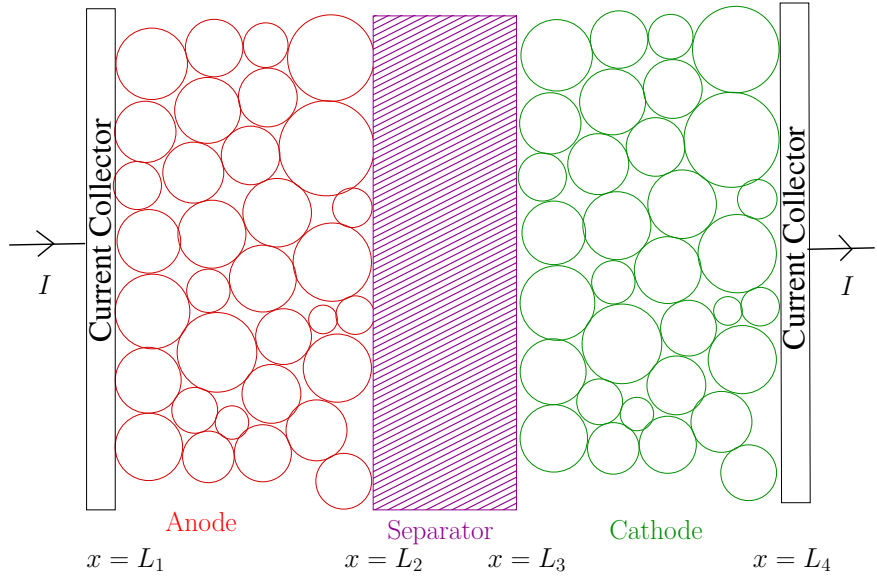}
\caption{Schematic of a planar Li-ion cell}
\label{schematic}
\end{center}
\end{figure}

In this section the Doyle Fuller Newman (DFN) model for lithium ion transport in a planar Li-ion cell (battery) comprised of an anode and a cathode separated by a porous spacer (as illustrated in figure \ref{schematic}) as first set out in  \cite{arora00,doyle93,fuller94b,fuller94,NewmanBook} is reprised below. The anode and cathode are both comprised of tightly packed spherical electrode particles, radii $R_a(x)$ and $R_c(x)$, respectively, and the interstices between these particles are filled with a finely structured porous  binder (treated with conductivity enhancing additives) which is filled with a lithium electrolyte. Here we consider a cell in which:
\be
\begin{array}{lc}
\mbox{The anode occupies} & L_1<x<L_2,\\
\mbox{the separator occupies} & L_2<x<L_3,\\
\mbox{the cathode occupies} & L_3<x<L_4.
\end{array} \label{cell-geom}
\ee
as illustrated in figure \ref{schematic}.
The model variables are tabulated and described in Table \ref{table1} while   the parameters and input functions are listed in Table \ref{table2}.
\begin{table}[ht]
\centering
\begin{tabular}{|c|c|c|}
\hline \textbf{Variable}  & \textbf{Description}  & \textbf{Units} \\
 &  &  \\
\hline $x$ & Distance across cell & m \\
\hline $t$ & Time & s \\
\hline $r$ & Radial distance from centre of electrode particle & m\\
\hline $c$ & Concentration of Li$^+$ ions in electrolyte & mol\,m$^{-3}$\\
\hline $\avvqn$ & Averaged flux of negative counterions in electrolyte & mol\,m$^{-2}$s$^{-1}$\\
\hline $\avvqp$ & Averaged flux of Li$^+$ ions in electrolyte & mol\,m$^{-2}$s$^{-1}$\\
\hline $\vph$ & Electric potential w.r.t. lithium electrode in electrolyte & V \\
\hline $\avvj$ & Averaged electrolyte current density  &  A\,m$^{-2}$  \\
\hline $\avjn$ & Component of current density on electrode particle surface & \\
                   & in direction of outward normal to particle & A\,m$^{-2}$  \\
\hline $j_a$ & Averaged current density in solid part of anode &  A\,m$^{-2}$ \\
\hline $j_c$ & Averaged current density in solid part of cathode &  A\,m$^{-2}$ \\
\hline $\Phi_a$ & Electric potential of anode (as function of position) & V \\
\hline $\Phi_c$ & Electric potential of cathode (as function of position) & V \\
\hline $c_a$ & Li$^+$ concentration in anode particles & mol\,m$^{-3}$\\
\hline $c_c$ & Li$^+$ concentration in cathode particles & mol\,m$^{-3}$\\
\hline $\eta_a$ & Overpotential between electrolyte and anode particles & V\\
\hline $\eta_c$ & Overpotential between electrolyte and cathode particles & V\\
\hline $V(t)$ & Potential drop across device & V\\
\hline
\end{tabular}
\caption{Variables used in the DFN model}
\label{table1}
\end{table}

\begin{table}[ht]
\centering
\begin{tabular}{|c|c|c|}
\hline \textbf{Param./}  & \textbf{Description}  & \textbf{Value and Units} \\
 \textbf{Ftn.}  &  &  \\
\hline $T$ & Absolute temperature &  K  \\
\hline $F$ & Faraday's constant &  A\,s\,mol$^{-1}$  \\
\hline $R$ & Universal gas constant & J\,K$^{-1}$mol$^{-1}$  \\
\hline ${\cal B}(x)$ & Inverse McMullin number & dim'less \\
\hline $\epsilon_l(x)$ & Volume fraction of electrolyte as function of position & dim'less \\
\hline $\bet(x)$ & Contact surf. area particles with electrolyte per unit vol. electrode &  m$^{-1}$ \\
\hline $\tp$ & Electrolyte transference number & dim'less \\
\hline $\De(c)$ & Ionic diffusivity of electrolyte  & m$^2$  s$^{-1}$ \\
\hline $\kappa(c)$ & Electrolyte conductivity  & A m$^{-1}$V$^{-1}$ \\
\hline $\sigma_a(x)$ & Anode conductivity as function of position & A m$^{-1}$V$^{-1}$ \\
\hline $\sigma_c(x)$ & Cathode conductivity as function of position & A m$^{-1}$V$^{-1}$\\
\hline $R_a(x)$ & Radius of anode particles & m \\
\hline $R_c(x)$ & Radius of cathode particles & m \\
\hline $\cam$ & Max. lithium concentration in anode particles & mol\,m$^{-3}$\\
\hline $\ccm$ & Max. lithium concentration in cathode particles & mol\,m$^{-3}$\\
\hline $k_a$ & Butler-Volmer constant in anode & mol$^{-1/2}$m$^{5/2}$s$^{-1}$ \\
\hline $k_c$ & Butler-Volmer constant in cathode & mol$^{-1/2}$m$^{5/2}$s$^{-1}$\\
\hline $\ueqa(c_a)$ & Open-circuit potenital: function of Li$^+$ concn. in anode & V\\
\hline $\ueqc(c_c)$ & Open-circuit potential: function of Li$^+$ concn. in cathode & V\\
\hline $D_a(c_a)$ & Li$^+$ diffusivity in anode particles: function of Li$^+$ concn. & m$^2$ s$^{-1}$ \\
\hline $D_c(c_c)$ & Li$^+$ diffusivity in cathode particles: function of Li$^+$ concn. & m$^2$  s$^{-1}$ \\
\hline $\mu_e(c)$ & Chemical potential of the electrolyte & J \, mol$^{-1}$ \\
\hline  $I(t)$  & Current flow through the cell & A\\
\hline $A$ & Area of cell & m$^2$\\
\hline
\end{tabular}
\caption{User specified functions and parameters in the DFN model}
\label{table2}
\end{table}

\subsection{The model equations}
The model naturally divides between spatially one-dimensional macroscopic equations which describe lithium transport and current flows on the scale of the whole cell, in the region $L_1<x<L_4$, and microscopic equations which describe lithium transport inside individual electrode particles. Since the particles are assumed to be spherical the microscopic transport equations are also one dimensional, but this time the spatial variable $r$ is the distance from the centre of a particle (here $0<r<R_a(x)$ within the anode and $0<r<R_c(x)$ within the cathode). The resulting model is thus two-dimensional in space.

\paragraph{Macroscopic Equations}
We start by writing down the macroscopic equations for lithium transport and current flow within the electrolyte. We denote $c$ as the lithium ion concentration,  $\avjn$ as current flux per unit surface area of electrode particle, and $\vph$ as the potential in the electrolyte, measured with respect to a lithium electrode (such that $\bar{\mu}_p=F \vph$, where $\bar{\mu}_p$ is the electrochemical potential of Li$^+$ ions in the electrolyte). The macroscopic electrolyte equations follow and are
\be
\begin{array}{cc}
\ds \ev \frac{\dd c}{\dd t}+ \frac{\dd}{\dd x} \avvqp= \frac{\bet}{F} \avjn,
&  \ds \avvqp=-\De(c) \cd  \frac{\dd c}{\dd x} + \tn \frac{\avvj}{F},  \\*[4mm]
\ds \frac{\dd \avvj}{\dd x}= \bet \avjn, 
& \ds \avvj= -\kappa(c) \cd \left( \frac{\dd \vph}{\dd x} - \frac{2}{F} (1-\tn)  \frac{d \mu_e}{d c}  \frac{\dd c}{\dd x} \right),
\end{array} \label{dfn1}
\ee
where $\avvqp$, $\avvj$ and the flux of lithium ions and current density in the electrolyte, respectively, averaged over the porous binder structure filling the interstices between the electrode particles.  In addition $\ev$ is the volume fraction of electrolyte, $\bet(x)$ is the surface area of electrode particles per unit volume, and the factor $\cd(x)$ is a dimensionless geometric factor that arises from the averaging (see, for example, \cite{richardson12,battrev}) which can be thought of as the inverse McMullin number. Note also that the final term in the equation for $\avvj$ involves the derivative of $\mu_e(c)$, the chemical potential of the electrolyte. This is not the usual way of expressing this term but it is the expression for the current that is obtained when the electrolyte equations are derived directly from the Stefan-Maxwell equations, see \cite{battrev}. \ma{Moreover it has been shown, in \cite{plett15} (on p.104), that $d \mu_e/dc=(RT/c) \left( 1+ {\dd \log f_\pm}/{\dd \log c} \right)$ from which it is readily seen that (\ref{dfn1}d) is equivalent to the expression for the averaged current density more commonly found in the literature.}

Macroscopic equations also need to be specified for the current flow in both the solid part anode and cathode matrices (\ie through the electrode particles and the conductively enhanced binder material). It is usual to assume that the averaged current density in both anode and cathode ($j_a$ and $j_c$, respectively) obey Ohm's law in which the currents are driven by gradients in the electrode potentials $\Phi_a$ (in the anode) and $\Phi_c$ (in the cathode), as given below in \eqref{dfn2}-\eqref{dfn5}. The final part of the macroscopic model that requires specification is the current flow  (current density $\avjn$) across the surfaces of the electrode particles into the electrolyte. This is usually specified in terms of a Butler-Volmer relation, as given below in \eqref{dfn6}-\eqref{dfn7}. 
\be
\frac{\partial j_a}{\partial x}=-\bet(x) \avjn,\quad 
j_a=-{\sigma_a}\frac{\partial \Phi_a}{\partial x} \quad \mbox{in} \quad L_1< x<L_2,~~~~~~~~~~~~~ \label{dfn2}\\
  j_a|_{x=L_1} = \frac{I(t)}{A}, \qquad j_a|_{x=L_2} = 0, ~~~~~~~~~~~~~~~~~~~~~~~~~~~~~~~~ \label{dfn3}\\
\frac{\partial j_c}{\partial x}=-\bet(x) \avjn,\quad 
j_c=-{\sigma_c}\frac{\partial \Phi_c}{\partial x} \quad \mbox{in} \quad L_3< x<L_4,~~~~~~~~~~~~~ \label{dfn4} \\
j_c|_{x=L_3} = 0, \qquad  j_c|_{x=L_4} = \frac{I(t)}{A}.~~~~~~~~~~~~~~~~~~~~~~~~~~~~~~~~~\label{dfn5} \\
\avjn=\left\{\begin{array}{ccc}
\ds 2 F k_a c^{1/2} \left( c_a \rvert _{r=R_a(x)} \right)^{1/2} \left( \cam -c_a \rvert _{r=R_a(x)}\right)^{1/2} \sinh\left(\frac{F \eta_a}{2RT}\right) &  \mbox{in} & L_1 \leq x< L_2,    \\
0 & \mbox{in} & L_2 < x<L_3,~~~ \\
\ds 2 F k_c c^{1/2} \left( c_c \rvert _{r=R_c(x)} \right)^{1/2} \left( \ccm -c_c \rvert _{r=R_c(x)}\right)^{1/2} \sinh\left(\frac{F\eta_c}{2RT}\right) &  \mbox{in} & L_3 \leq x< L_4,  \end{array} \right.                                                                    \label{dfn6} \\
 \eta_a=\Phi_a-\vph-\ueqa(c_a|_{r=R_a(x)}), \qquad \eta_c=\Phi_c-\vph-\ueqc(c_c|_{r=R_c(x)}).~~~~~~~~~~~~~\label{dfn7} 
\ee
Here $\sigma_a$  and $\sigma_c$ are the conductivities in the solid part of the anode and the cathode, respectively.

\paragraph{Microscopic equations and boundary conditions}
Transport of lithium in both anode and cathode particles occurs through a nonlinear diffusion process, and given the spherical symmetry of the particles this may be modelled by the following diffusion equations, in which $r$ measures distance from the centre of the particle
\be
 \left. \begin{array}{l} \ds \frac{\partial c_a}{\partial t}=\frac{1}{r^{2}}\frac{\partial }{\partial r}\left( r^{2} D_a (c_a)\frac{\partial c_a}{\partial r}\right) \quad \mbox{in} \quad 0< r<R_a(x). \\
\ds c_a \,\, \mbox{bounded} \,\, \mbox{on} \,\, r=0, \qquad -D_a(c_a) \frac{\partial c_a}{\partial r}\bigg\rvert_{r= R_a(x)}=\frac{\avjn}{F} 
\end{array} \right\}\ \  \mbox{in} \ \  L_1< x<L_2,  \label{dfn8} \\
 \left. \begin{array}{l} \ds \frac{\partial c_c}{\partial t}=\frac{1}{r^{2}}\frac{\partial }{\partial r}\left( r^{2} D_c (c_c)\frac{\partial c_c}{\partial r}\right) \quad \mbox{in} \quad 0< r<R_c(x)\\
\ds c_c \,\, \mbox{bounded} \,\, \mbox{on} \,\, r=0, \qquad -D_c(c_c) \frac{\partial c_c}{\partial r}\bigg\rvert_{r= R_c(x)}=\frac{\avjn}{F} 
\end{array} \right\}\ \  \mbox{in} \ \  L_3< x<L_4.  \label{dfn9} 
\ee
Note that the boundary conditions on the surfaces of these particles relate the surface lithium ion flux to the reaction current $\avjn$ .

\paragraph{The full cell potential.} 
The results of solution to the full cell DFN model (with appropriate initial conditions) and a specified galvanostatic current $I(t)$ can be used to compute the potentials of the anode and cathode current collectors $V_a$ and $V_c$, respectively via the relations
\be \label{fc50}
V_a(t)= \Phi_a \big\rvert_{x=L_1}, \qquad V_c(t)= \Phi_c \big\rvert_{x=L_4}. \label{dfn10} 
\ee
and hence the potential drop $V(t)$ occurring across the cell
\be
V(t)=V_c(t)-V_a(t). \label{dfn11} 
\ee

\paragraph{An alternative formulation of the electrolyte equations.}
We note that we could re-express equations (\ref{dfn1}a-b) in terms of a conservation law for negative counterions (as opposed to one for positive lithium ions). On denoting the averaged flux of the negative counterions by $\avvqn$ this reads
\be
\begin{array}{cc}
\ds \ev \frac{\dd c}{\dd t}+ \frac{\dd}{\dd x} \avvqn= 0,
&  \ds \avvqn=-\De(c) \cd  \frac{\dd c}{\dd x} -(1- \tn) \frac{\avvj}{F},
\end{array}  \label{biden2}  \label{dfn12} 
\ee
and can readily be verified to be equivalent to (\ref{dfn1}a-b) on making use of (\ref{dfn1}c).


\section{Energy Conservation Law for the DFN model \label{encons}}

\paragraph{Chemical energy stored in the cell.}
The Gibbs free energy of the cell is predominantly stored in the electrode materials but, in a working device, there is also a minor contribution from concentration gradients in the electrolyte. The Gibbs free energy density of the lithium ions stored in the anode material $\G_a(c_a)$ and the cathode material $\G_c(c_c)$ are given in terms of the open circuit voltages by the expressions
\be
\G_a(c_a)=-\int F U_{eq,a}(c_a) d c_a \quad \mbox{and} \quad \G_c(c_c)=-\int F U_{eq,c}(c_c) d c_c, \label{gibbs_part}
\ee
while the Gibbs free energy density of the electrolyte $\G_e$ can be expressed in terms of the electrolyte chemical potential $\mu_e(c)$ as
\be
\G_e(c)= \int 2 \mu_e(c) dc.
\ee
Here the factor of 2 is to account for the fact that both the distribution of Li$^+$ ions and negative counterions (which are at equal concentration $c$) contribute to the chemical energy. The total Gibbs energy stored in the cell is thus the sum of the integral of $\G_a$ over all the anode particles, and the the integral of $\G_c$ over all the cathode particles, and of $\G_e$ through the electrolyte. On denoting  the chemical energy, per unit area of cell, as $\gibbs$ we arrive at the following expression:
\be
\gibbs(t)= \int_{L_1}^{L_2} \N_a(x) \left(\int_0^{R_a(x)} 4 \pi r^2 \G_a(c_a) dr\right) dx+ \int_{L_3}^{L_4} \N_c(x) \left(\int_0^{R_c(x)} 4 \pi r^2 \G_c(c_c) dr\right) dx \non \\
 + \int_{L_1}^{L_4}\ev \G_e(c) dx.~~~~~~~~~~~~ \label{echem}
\ee
where $\N_a(x)$ and $\N_c(x)$ are the number of electrode particles per unit volume in the anode and cathode, respectively, and $\ev(x)$ is the volume fraction of the electrolyte.

\paragraph{The energy conservation equation.}

\ma{In what follows we write down a conservation law for the total Gibbs free energy within an isothermal cell, held at constant temperature $T$. This law shows that, in a cell of area $A$, the loss of Gibbs energy $A\, \gibbs$ can be equated to the useful work extracted from the cell, in the form of a current flow $I$ across a potential difference $V$, and a number of energy loss terms. As we shall demonstrate these energy loss terms can be identified with irreversible heating. The energy conservation equation is a direct consequence of the DFN model and, in \S \ref{deriv}, a mathematically rigorous derivation is made. The conservation law states that minus the rate of change of Gibbs energy in the cell is equal to the power drawn from the device (in the form of current), plus the sum of a number of irreversible heat production rates (representing the losses occurring in the constituent parts of the cell) and can be written in the form of the equation
\be
-A\frac{d \gibbs}{d t} = {I V}+ A \left(\helec+ \hama+\hohma+\hopa+\hamc+\hohmc+\hopc\right).~~~~ \label{econs-fin}
\ee
More specifically the terms on the right-hand side of this relation are (from left to right) the power extracted by the circuit $IV$, and the heat generated by: (I) dissipative effects in the electrolyte $A \helec$, (II) the heat of mixing in the anode particles $A \hama$, (III) Ohmic dissipation in the solid parts of the anode $A \hohma$, (IV) dissipation arising from the current flow across the overpotential drop at the surfaces of the anode particles $A\hopa$ (also termed the polarisation loss), (V)   the heat of mixing in the cathode particles $A \hamc$, (VI) Ohmic dissipation in the solid parts of the cathode $A \hohmc$, (VII) the polarisation losses at the surfaces of the cathode particles $A\hopc$. The heat dissipation terms are calculated from the solution to the DFN model \eqref{dfn1}-\eqref{dfn11}, as shown in \S \ref{deriv}, via the following integral expressions:}
\be
\ds \helec=\int_{L_1}^{L_4}\left( 2 \cd \De(c) \frac{d \mu_e}{d c} \left(\frac{\dd c}{\dd x} \right)^2+  \frac{1}{\kappa(c) \cd} \avvj^2 \right) dx, \label{econs1} \\*[4mm]
\ds \hama= -4 \pi F \int_{L_1}^{L_2} \N_a(x) \left(\int_0^{R_a(x)} D_a(c_a) \left( \frac{\dd c_a}{\dd r} \right)^2  \frac{d U_{eq,a}}{d c_a}  \, r^2 dr \right)  dx, \label{econs2}\\*[4mm]
\ds \hamc= -4 \pi F \int_{L_3}^{L_4} \N_c(x)\left(\int_0^{R_c(x)} D_c(c_c) \left( \frac{\dd c_c}{\dd r} \right)^2  \frac{d U_{eq,c}}{d c_c}  \, r^2 dr \right) dx, \label{econs3}\\*[4mm]
\begin{array}{ll}
\ds \hohma=\int_{L_1}^{L_2} \sigma_a \left( \frac{\dd \Phi_a}{\dd x} \right)^2 dx, & \ds \hopa=\int_{L_1}^{L_2}\bet(x)\eta_a \avjn  dx,  \\*[4mm]
\ds \hohmc= \int_{L_3}^{L_4}\sigma_c \left( \frac{\dd \Phi_c}{\dd x} \right)^2 dx, & \ds \hopc=\int_{L_3}^{L_4}\bet(x)\eta_c \avjn dx.
\end{array} \label{econs4}
\ee
Here the number densities of electrode particles, per unit volume, in the anode and cathode are related to the radii of the electrode particles and the B.E.T. surface area $b_{et}(x)$ (surface area of particle per unit volume of electrode) via the identities
\be
\N_a(x)=\frac{\bet(x)}{4 \pi R_a^2(x)} \quad \mbox{in} \quad L_1<x<L_2, \qquad 
\N_c(x)=\frac{\bet(x)}{4 \pi R_c^2(x)} \quad \mbox{in} \quad L_3<x<L_4. \label{econs5}
\ee
Note, that despite the minus signs appearing in the expressions for $\hama$ and $\hamc$ these quantities are expected to be positive because both $U_{eq,a}'(c_a)$  and $U_{eq,c}'(c_c)$ are negative.

{\subsection{Heat production within the cell.} The energy conservation law can be used to determine the heat produced by the cell provided we relax the isothermal constraint and consider instead the cell is at a uniform, time-dependent temperature $T(t)$; the assumption that temperature is spatially independent is reasonable because single cells are extremely thin.  Since the open circuit voltages $U_{eq,a}$ and $U_{eq,c}$ are usually weak functions of temperature, we write $\gibbs=\gibbs(t,T)$, so that in this instance
\be
\frac{d \gibbs}{d t}= -S \frac{d T}{d t}+\left.\frac{\dd \gibbs}{\dd t} \right|_T,
\ee
where $S$ is the entropy of the cell (per unit area) and we make use of the identity ${\dd \gibbs}/{\dd T}=-S$. In addition we denote the time derivative of the Gibbs energy taken at constant temperature by $\left.{\dd \gibbs}/{\dd t} \right|_T$  and note that it can be identified as the total derivative of $\gibbs$ appearing in the isothermal conservation law \eqref{econs-fin}. The heat production of the cell can be calculated from the enthalpy (per unit area) $H$, which is related to the Gibbs energy  by the standard formula $H=G+TS$. It follows that 
\be
\frac{d H}{d t}= T \frac{d S}{d t} +\left.\frac{\dd \gibbs}{\dd t} \right|_T.
\ee
Furthermore, the rate of change of enthalpy is given in terms $\hrev+\hirr$, the rate of heat dissipation (per unit area), and $\wdot$, the rate of work done by the cell (per unit area), by the formula
\bes
\frac{d H}{d t}=-\hrev-\hirr - \wdot, \quad \mbox{where} \quad \hrev=-T \frac{d S}{d t}.
\ees
Here the total rate of heat dissipation is split into a reversible heating term, $\hrev$, and an irreversible one, $\hirr$. On substituting for $\left.{\dd \gibbs}/{\dd t} \right|_T$ from the isothermal conservation law \eqref{econs-fin} we can identify
\be
\hirr=\helec+ \hama+\hohma+\hopa+\hamc+\hohmc+\hopc, \quad \mbox{and} \quad \wdot=\frac{IV}{A}.
\ee
In order to compute an expression for the rate of reversible heating $\hrev$ we use the facts that $S=-{\dd \gibbs}/{\dd T}$, and that it is only significant within the electrode particles, to obtain the expression
\be
\hrev=-F T \frac{d}{dt}\left[\int_{L_1}^{L_2} \N_a(x) \left(\int_0^{R_a(x)} 4 \pi r^2  \left(\int \frac{\dd U_{eq,a}}{\dd T}(c_a) d c_a \right)  dr\right) dx \right. \non \\
\left.+ \int_{L_3}^{L_4} \N_c(x) \left(\int_0^{R_c(x)} 4 \pi r^2 \left(\int \frac{\dd U_{eq,c}}{\dd T}(c_c) d c_c \right) dr\right) dx\right], \label{rev_heat}
\ee
Here we have made use of the expressions for the Gibbs free energy densities in the electrode particles found in \eqref{gibbs_part} and the chemical energy (per unit area) $G$ found in \eqref{echem}.
Notably this is not the expression that is typically used to compute the reversible heating. However, where we make the approximation that the particles discharge uniformly, so that $c_a \approx c_a(x,t)$ and $c_c \approx c_c(x,t)$ (\ie there is no radial dependence of the lithium concentrations within the electrode particles), it can be shown that
\be
\hrev \approx T\left[\int_{L_1}^{L_2} \bet \avjn \frac{\dd U_{eq,a}}{\dd T}  dx + \int_{L_3}^{L_4} \bet \avjn \frac{\dd U_{eq,c}}{\dd T} dx \right], 
\ee
by equating the rate of change of the lithium concentrations within particles to the flux of lithium being transported into/out of the particles by the reaction currents on their surfaces $\avjn$. This is precisely the expression that is normally used to compute the contribution of the rate of reversible heating, see for example \cite{lai15,du17}. Furthermore, this expression is likely to be fairly accurate except where the discharge is sufficiently aggressive to cause significant gradients in lithium concentration within the electrode particles.  We note that data for ${\dd U_{eq}}/{\dd T}$, as a function of $c_s$, can be found for most electrode materials. For example, \cite{du17} gives data for graphite and LFP, while \cite{farag17} gives it for graphite and NMC.}

\section{Results \label{res}}

\begin{figure} \centering
\includegraphics[width=0.7\textwidth]{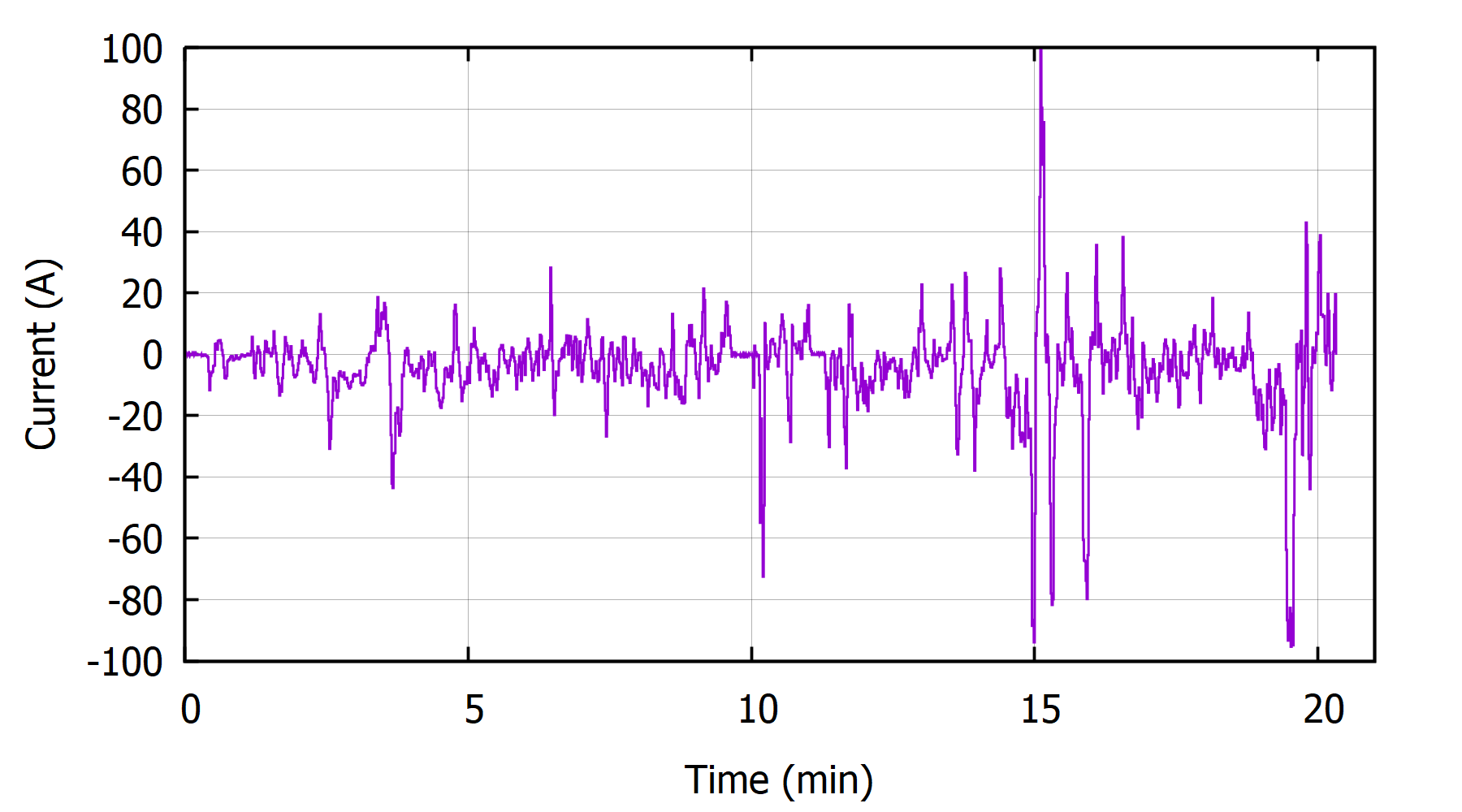}\\
\caption{A single pattern of the drive cycle current.}
\label{fig:DC_current}
\end{figure}

\begin{figure} \centering
\includegraphics[width=0.8\textwidth]{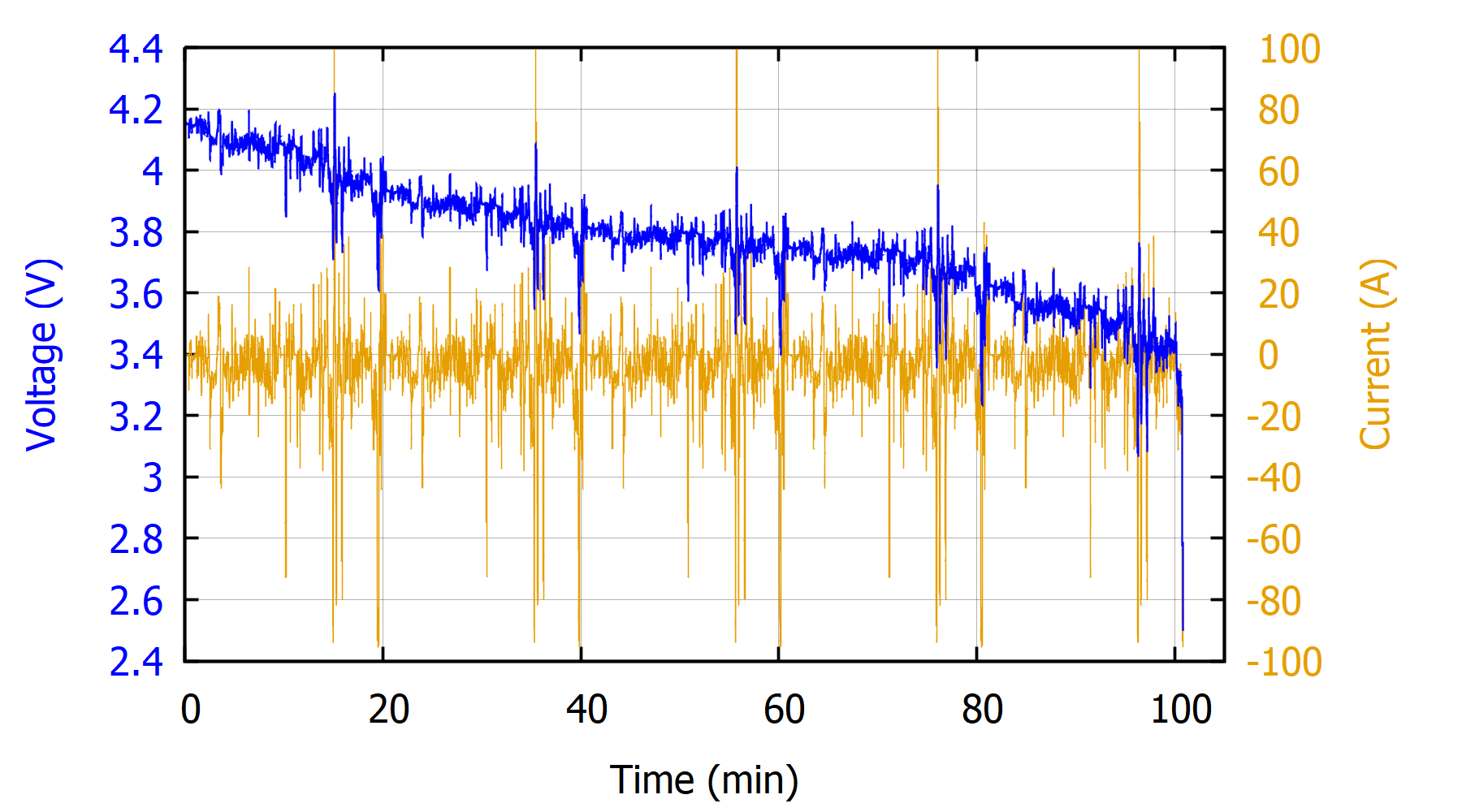}\\
\caption{Total voltage and periodically repeated drive cycle current versus time, a full discharge.}
\label{fig:DC_voltage}
\end{figure}

In this section compute solutions to the DFN model given in \eqref{dfn1}-\eqref{dfn11} with parameters taken from a study by Ecker and co-authors in \cite{Ecker2015a,ecker15}, on a 7.5 Ah cell produced by Kokam with a graphite anode and a Li(NiMnCo)O$_2$ cathode. We do this both for two constant discharge rates of 5C and 10C and for a drive cycle with current $I(t)$ reproduced in figure \ref{fig:DC_current} with the corresponding cell voltage (computed from the model) shown in figure \ref{fig:DC_voltage}. The point of this exercise is not only to validate the energy conservation equation \eqref{econs-fin}, and the auxillary equations \eqref{econs1}-\eqref{econs5}, but also to identify the sources of major energy loss in the cell under different operating conditions.

\begin{figure} \centering
\includegraphics[width=0.7\textwidth]{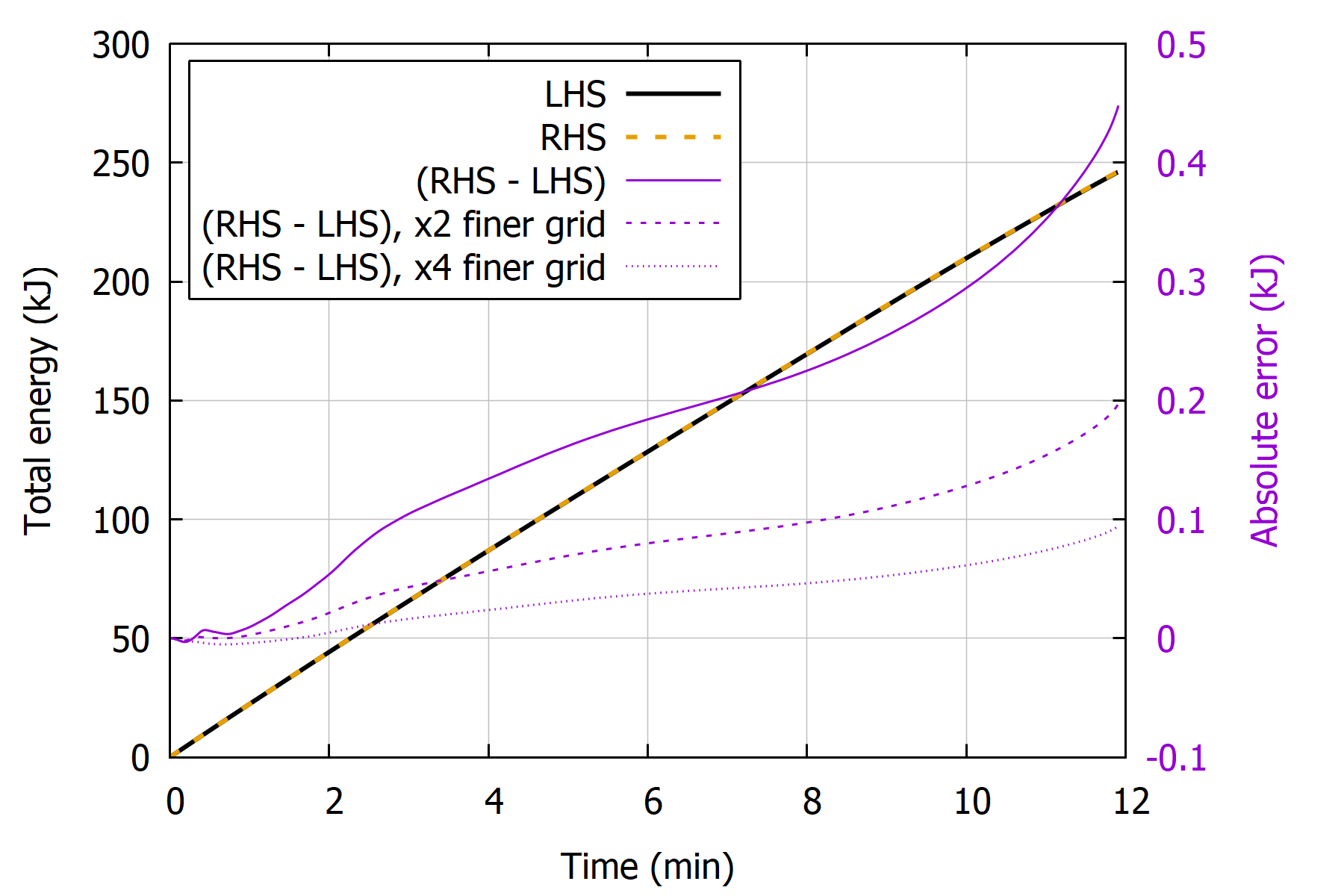}\\
\caption{Comparison of the LHS and the RHS of equation \eqref{econs-fin} numerically calculated from the DFN model \eqref{dfn1}-\eqref{dfn11} at constant current discharge rate 5C and integrated over discharge time.}
\label{fig:errors_5C}
\end{figure}

\begin{figure} \centering
\includegraphics[width=0.7\textwidth]{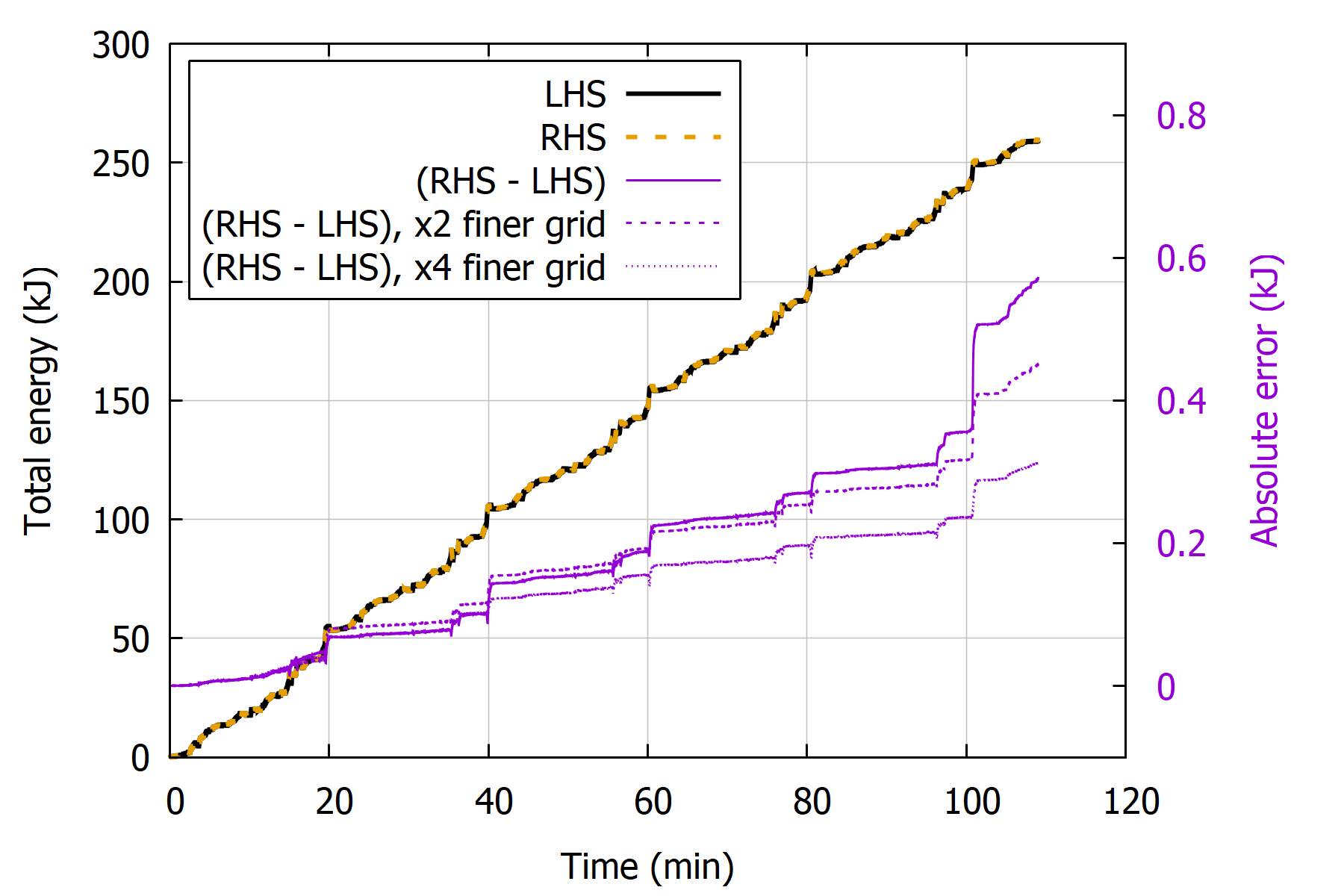}\\
\caption{Comparison of the LHS and the RHS of equation \eqref{econs-fin} numerically calculated from the DFN model \eqref{dfn1}-\eqref{dfn11} for the drive cycle and integrated over discharge time.}
\label{fig:errors_DC}
\end{figure}

\paragraph{Numerical Validation of the Energy Conservation Law.}
In figure \ref{fig:errors_5C} we validate the energy conservation law by computing integral of the the left-hand side of equation \eqref{econs-fin} over time, from the numerical solution  to the model,  at constant 5C discharge (black curve) and comparing this to integral of the right-hand side of equation \eqref{econs-fin} over time (also computed from the numerical solution). With the standard grid spacing the error is less than $0.2 \%$ (solid purple curve) but reduces as the accuracy of the numerical solution is improved further by choosing a finer grid resolution (dashed and dotted purple curves). An equivalent plot for the drive cycle is shown in figure \ref{fig:errors_DC}, and once again the error between computations of the left- and right-hand-sides of the \eqref{econs-fin} are attributable to discretisation error in the solution of the DFN model.

\begin{figure} \centering
\includegraphics[width=0.495\textwidth]{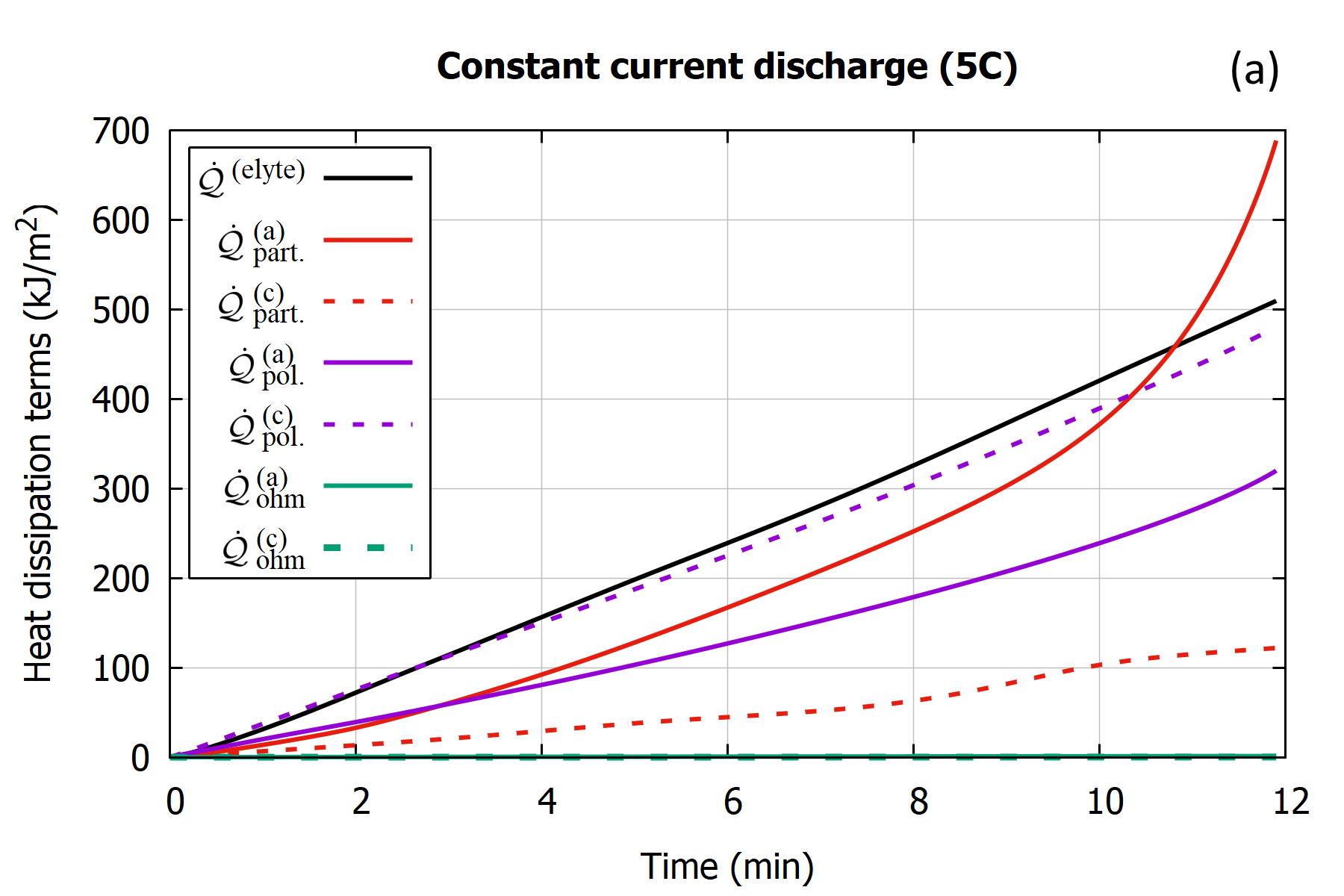}
\includegraphics[width=0.495\textwidth]{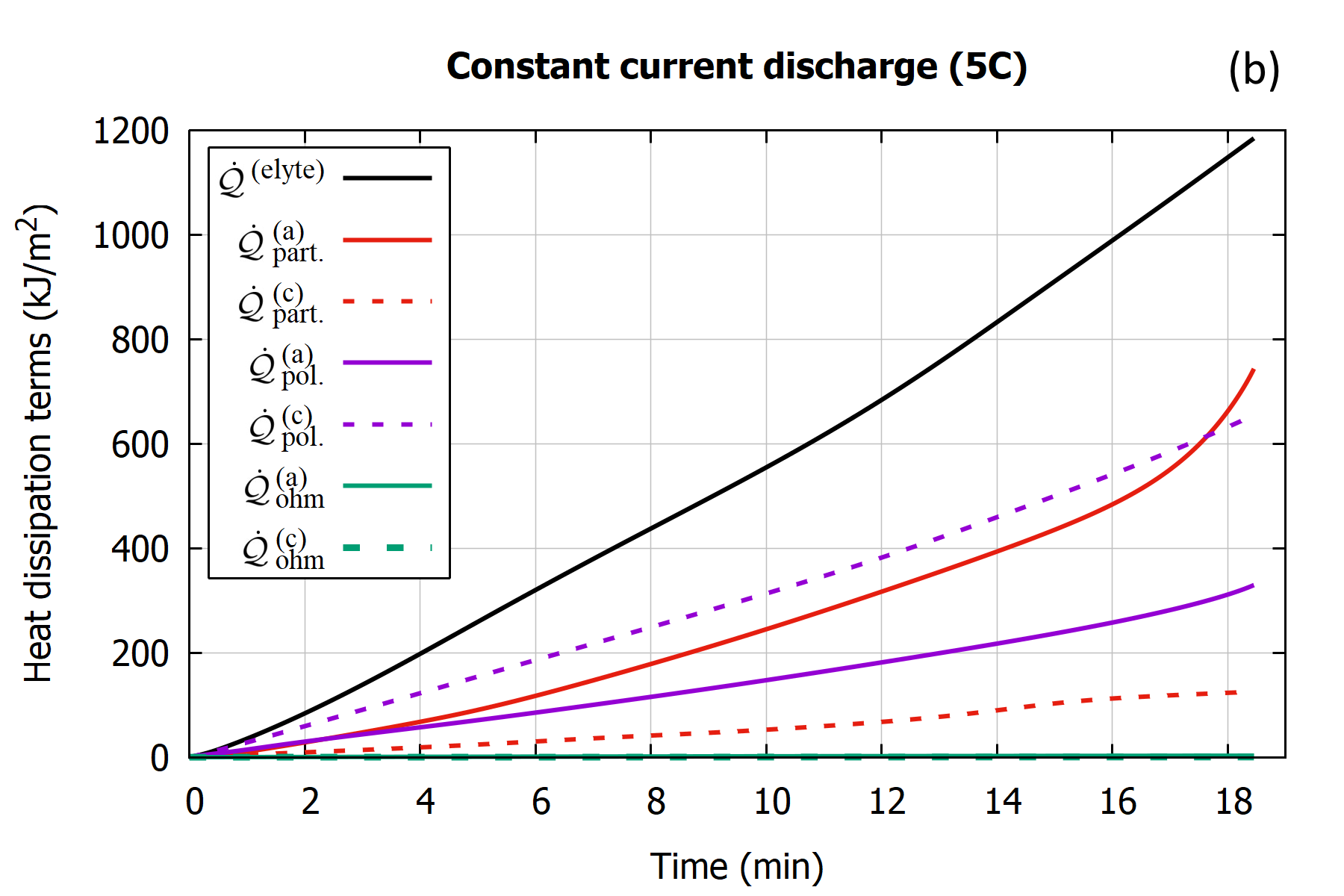}\\
\medskip
\includegraphics[width=0.495\textwidth]{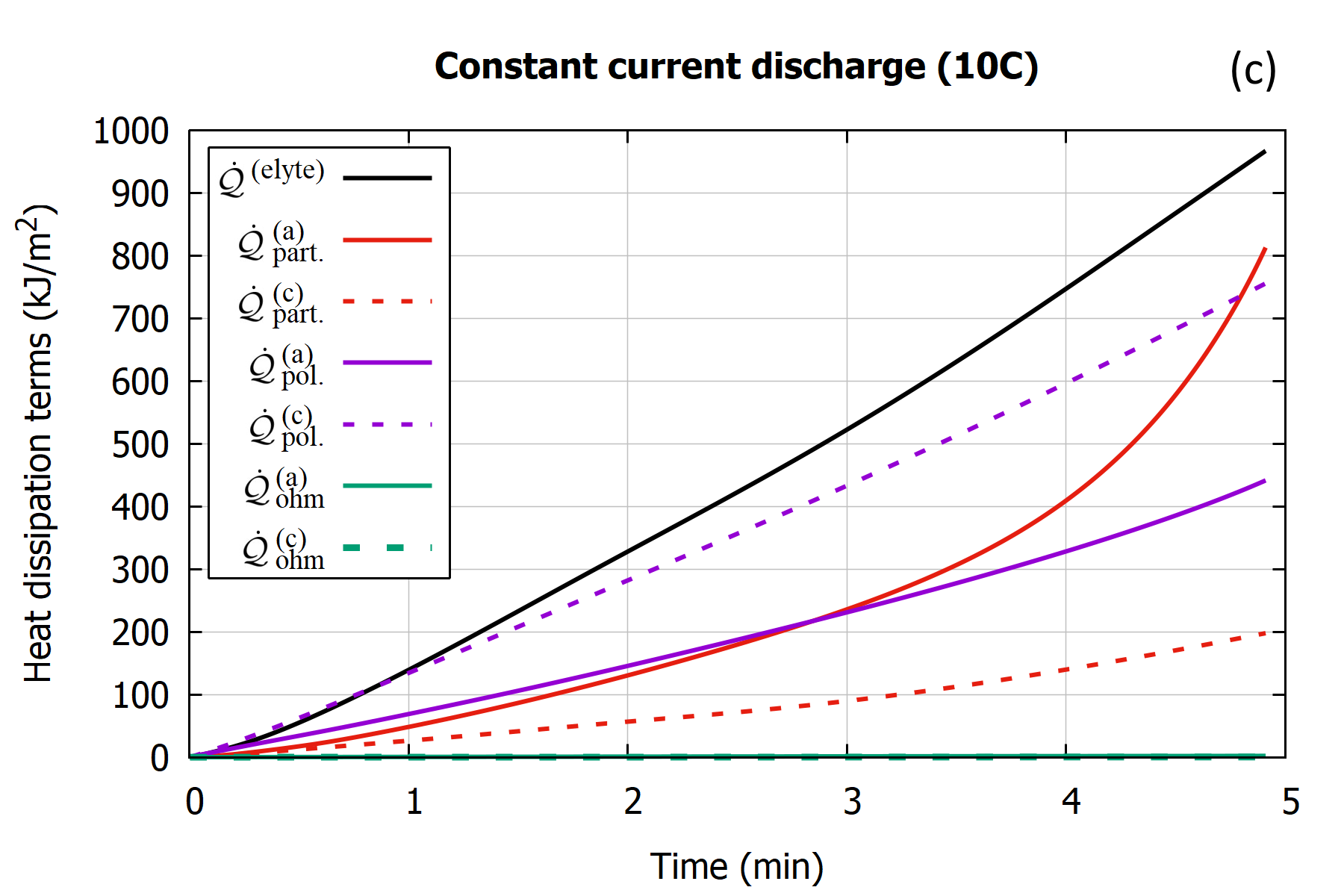}
\includegraphics[width=0.495\textwidth]{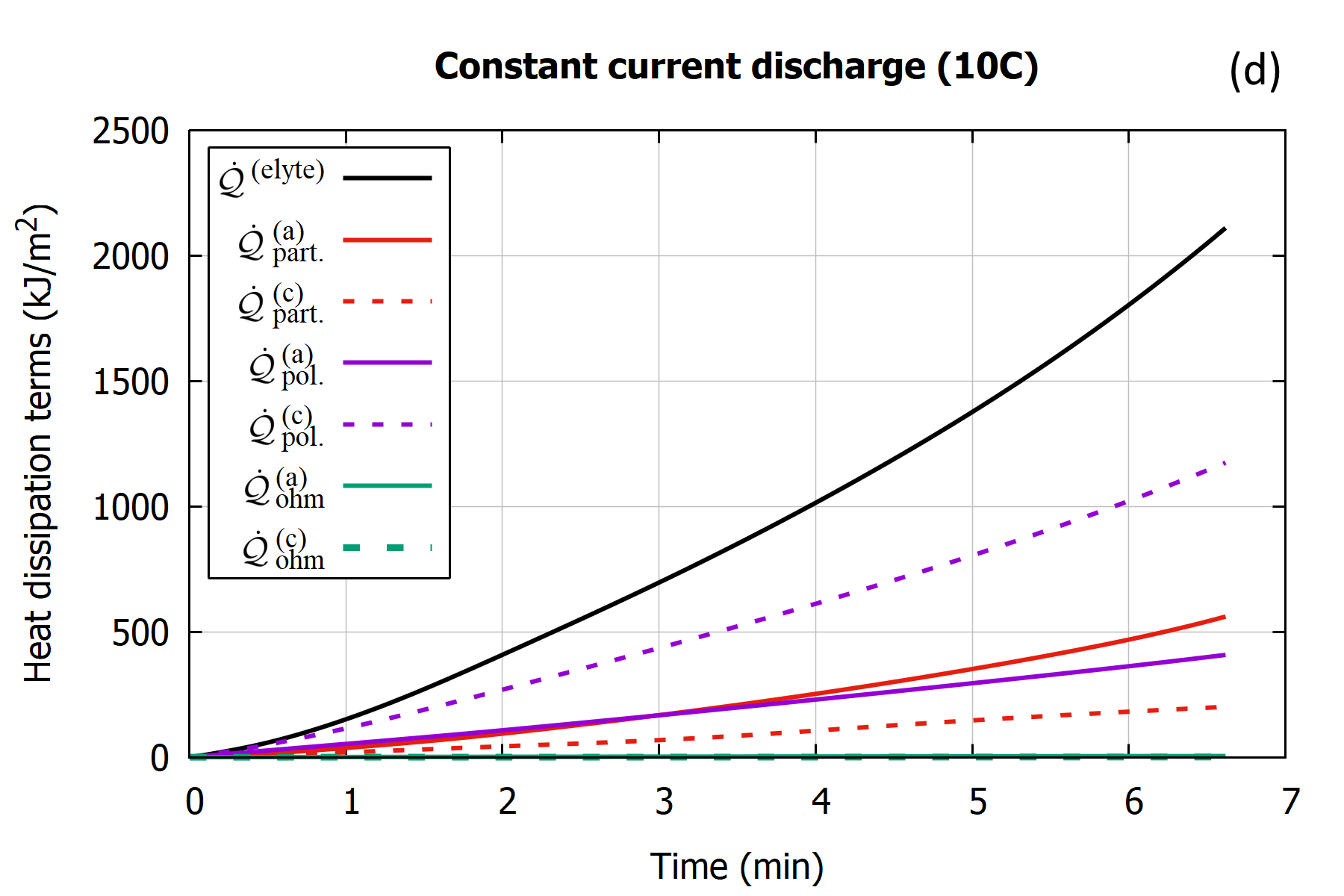}\\
\medskip
\includegraphics[width=0.495\textwidth]{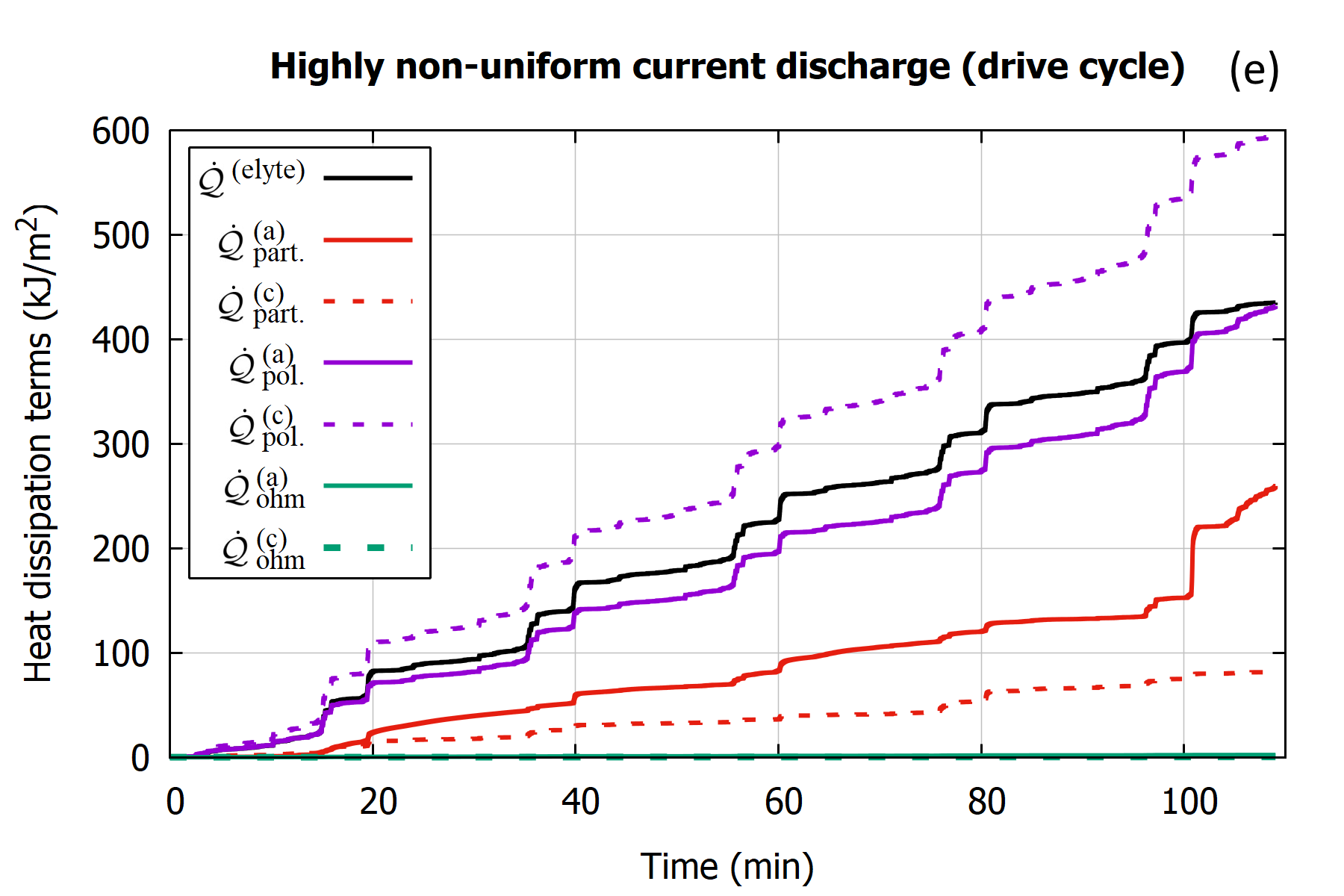}
\includegraphics[width=0.495\textwidth]{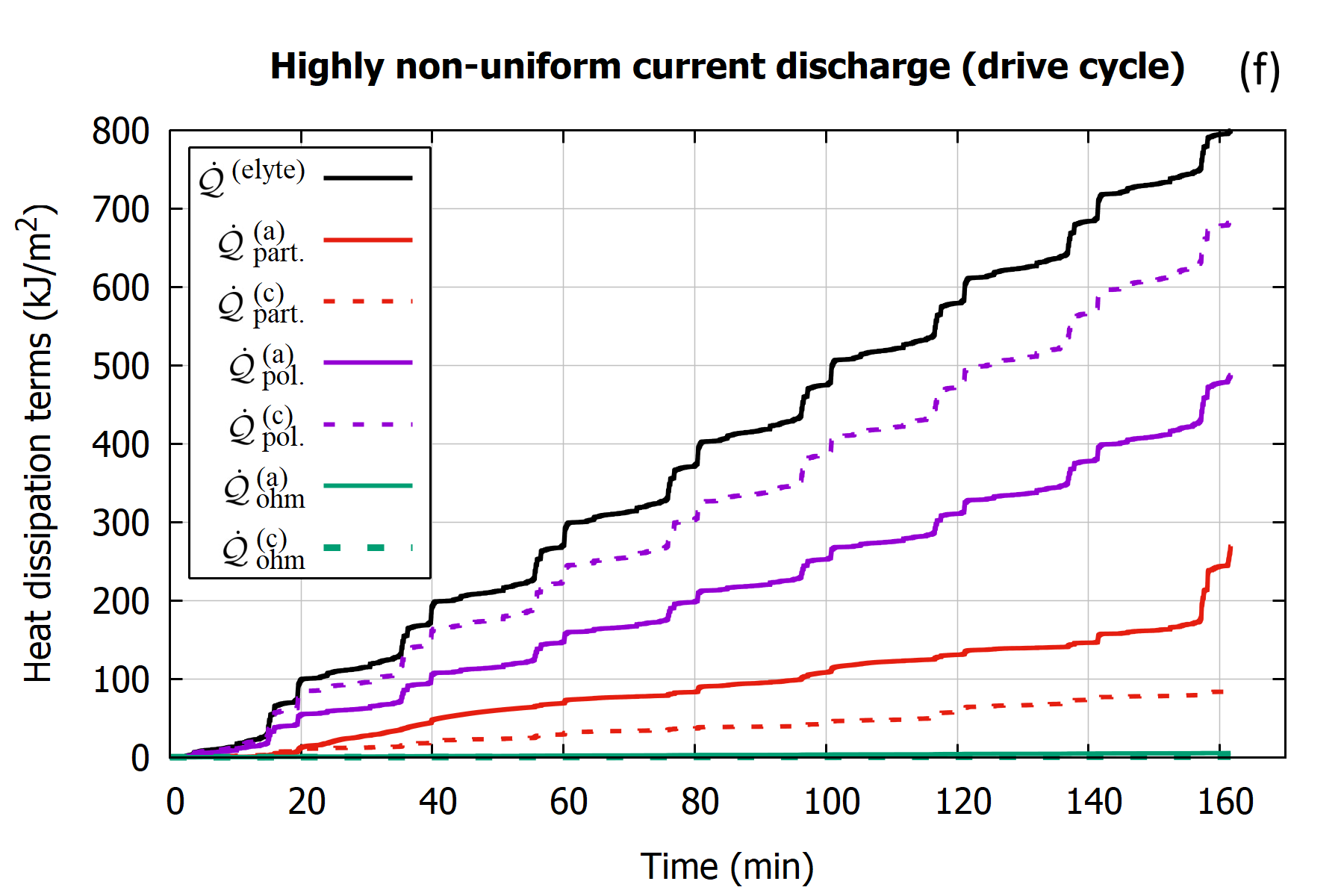}\\
\caption{Integrated heat dissipation terms \eqref{econs1}-\eqref{econs4} for a single constant current discharge at 5C rate (top figures), 10C discharge rate (middle figures), and for full discharge under periodically repeated drive cycle current shown in figure \ref{fig:DC_current} (bottom figures). The left column of figures corresponds to the standard electrode thickness according to \cite{ecker15}, the right column shows the heating terms for the increased thickness of both electrodes by a factor of 1.5.}
\label{fig:heat_terms}
\end{figure}

\paragraph{Use of the Energy Conservation Law in cell design.} 
In order to illustrate the possible use of the conservation law in cell design we consider the three scenarios outlined above (5C, 10C and drive cycle discharges) for both the standard 7.5 Ah Kokam cell and one with electrodes whose thickness has been increased by a factor of 1.5. In order to illustrate where the energy losses occur within the cell we split up the dissipation terms based on their location and plot the integral of each term over time in figure \ref{fig:heat_terms}. On the left-hand side of this figure we show the results for the standard cell while on the right-hand side we show those for the thicker cell. For the standard cell thickness (left) the integrated energy loss, at the end of the 5C cell discharge (12 mins), from the enthalpy of mixing term in the anode particles (solid red curve) is the dominant one. This in itself is remarkable as this term is neglected in nearly all the DFN treatments of energy loss that we are aware of. The other significant losses arise from the electrolyte (black) and polarisation across the interfaces of the cathode and anode particles (dashed and solid purple lines, respectively). The picture alters slightly for the 10C discharge, for which electrolyte losses become dominant but is very different for the drive cycle for which, perhaps unsurprisingly, the polarisation losses are considerably more important. Increasing the thickness of the electrodes (right) causes loss in the electrolyte to become much more significant, and while this is obviously a major limitation on cell performance for the constant current discharges it is not as significant for the drive cycle, for which electrolyte losses are still comparable to polarisation losses. Thus, while increasing electrode thickness in devices that are designed to undergo this drive cycle marginally increases the energy losses, it could reasonably be viewed as an acceptable downside that might be offset by the higher overall energy density afforded by thicker electrodes.


\section{Derivation of the Energy conservation equation \label{deriv}}

In order to derive the energy conservation from the underlying DFN model we consider conservation of Gibbs energy in the electrode particles (in \S \ref{epart}), the electrolyte (in \S \ref{elyte}) and the solid conductive parts of the electrodes (in \S \ref{solid}) before pulling together the pieces in \S \ref{whole} to obtain the final result.

\subsection{Energy conservation in an active material particle \label{epart}}
In both the anode and the cathode the lithium transport equations in the active material particles (\ie \eqref{dfn8} and \eqref{dfn9}) have the from
\be
\frac{\dd c_s}{\dd t}+\frac{1}{r^2} \frac{\dd}{\dd r} \left(r^2 N_s \right)=0 \quad  \mbox{in} \quad 0 \leq r<R_s(x), \qquad
\left. N_s \right|_{r=R_s(x)}=\frac{\avjn}{F},\label{li-cons}
\ee
where
\be
N_s=- D_s(c_s)\frac{\dd c_s}{\dd r}. \label{s-flux}
\ee 
is the flux of lithium ions in the radial direction and $s=a$ in the anode and $s=c$ in the cathode.

\paragraph{Chemical potential and Gibbs free energy density in the active material.} The chemical potential $\mu_s(c_s)$ in the active material as a function of lithium concentration is given by the relation
\be
\mu_s(c_s)=-F U_{eq,s} (c_s). \label{chempot_s}
\ee
where $U_{eq,s} (c_s)$ is the equilibrium potential of the material as a function of lithium concentration $c_s$. The Gibbs free energy density 
$\G_s(c_s)$ can then be calculated from the relation $\mu_s=d \G_s/d c_s$, from which we can deduce that 
\be
\G_s(c_s)= -F \int U_{eq,s} (c_s) d c_s.
\ee

\paragraph{A conservation equation for the chemical energy.} The local energy flux in the radial direction $N_{E,s}$ is found my multiplying the derivative of $\G_s$, with respect to concentration, by the particle flux $N_s$ such that
\be
N_{E,s}= \mu_s N_s, \label{organ}
\ee
where $N_s$ is given in \eqref{s-flux}. In order to derive a local energy conservation equation we take the derivative of $\G_s$ with respect to $t$
\bes
\frac{\dd \G_s}{\dd t}=\frac{d \G_s}{d c_s} \frac{\dd c_s}{\dd t}=\mu_s \frac{\dd c_s}{\dd t}
\ees
and on substituting for the time derivative of $c_s$ from \eqref{li-cons}, and rearranging, we obtain the energy conservation law
\be
\frac{\dd \G_s}{\dd t}+\frac{1}{r^2} \frac{\dd}{\dd r}  \left( r^2 \mu_s N_s \right) =N_s \frac{\dd \mu_s}{\dd r}. \label{energy_cons}
\ee
We can rewrite this in the following form to make it explicit that this is an energy conservation equation
\be
\frac{\dd \G_s}{\dd t}+\frac{1}{r^2} \frac{\dd}{\dd r}  \left( r^2 N_{E,s} \right) = -\Omega \quad \mbox{where} \quad \Omega=-N_s \frac{\dd \mu_s}{\dd r}, \label{energy_cons2}
\ee
by identifying $\mu_s N_s$ as the energy flux (in the radial direction) $N_{E,s}$  and denoting the rate of loss of chemical energy, per unit volume, as $\Omega$. This chemical energy is transferred into heat energy and so $\Omega$ is also the rate of heat energy production, per unit volume. 

\paragraph{An integral energy conservation law.}
The total chemical energy $\ecs$ stored in the electrode particle occupying the region $0<r<R_s$ is found by integrating the Gibbs free energy density over the particle so that
\be
\ecs=\int_0^{R_s(x)} 4 \pi r^2 \G_s(c_s) dr. \label{ecs}
\ee
Note that the chemical potential is related to the Gibbs free energy via $\mu_s=d \G_s/d c_s$. 
On integrating \eqref{energy_cons2}, multiplied by $4 \pi r^2$, over the particle we obtain the integral form of the energy conservation law in the form of an evolution equation for the $\ecs$ (as defined in \eqref{ecs}), the total amount of chemical energy stored in the electrode particle,
\be
\frac{d \ecs}{d t}=  \left. -4 \pi R_s^2 \mu_s N_s \right|_{r=R_s} + 4 \pi \int_0^{R_s} r^2 N_s \frac{d \mu_s}{d c_s} \frac{\dd c_s}{d r} dr. \label{int-cons}
\ee
On substituting for $N_s|_{r=R_s}$ from (\ref{li-cons}b), for $\mu_s$ from \eqref{chempot_s} and for $N_s$ from \eqref{s-flux} this relation can be rewritten in the form
\be
\frac{d \ecs}{d t}= 4\pi R_s^2  (\avjn U_{eq,s}(c_s|_{r=R}) )-\omega^{\rm (heat)}_{\rm part,s}(x,t),~~~~~~~~~~~~~~~~~\label{encons1} \\
\mbox{where} \quad \omega^{\rm (heat)}_{\rm part,s}(x,t)=-4 \pi F\int_0^{R_s} D_s(c_s) \left( \frac{\dd c_s}{\dd r} \right)^2  \frac{d U_{eq,s}}{d c_s}  \, r^2 dr. \label{encons2} 
\ee
Equation \eqref{encons1} may be interpretted as saying that the rate of gain of chemical energy in the electrode particle is equal to the  rate of flow of energy into the particle $4\pi R_s^2  (j_n U_{eq,s}(c_s) )$ minus the rate of heat release in the particle $\omega^{\rm (heat)}_{\rm part,s}$. Notice that we would expect $\omega^{\rm (heat)}_{\rm part,s}$ to be a positive quantity because $U_{eq,s}'(c_s)$ is negative.


\subsection{An energy conservation law in the electrolyte \label{elyte}} 

{In Rao and Newman \cite{rao97} two methods are suggested to calculate the energy dissipation (to heat) in a lithium electrolyte. Both methods turn out to be ways of approximating the heat generated within the electrolyte. To quote directly from this work ``We notice that the local-heat-generation method has an inherent difference from the energy balance method. The former considers transport and kinetic phenomena within the cell while the latter uses a thermodynamic approach to the cell system. The assumptions used to reach the final heat effects are also different: the local-heat-generation methods neglects effects of any concentration gradient in the heat generation by electrical current (...), while the energy-balance method neglects mixing effects. These two assumptions are no doubt interrelated and present an interesting study for future study".}

{The purpose of this section is to derive the energy conservation law for an electrolyte filling a porous electrode and, as a by-product, deduce the irreversible energy dissipation to heat within such an electrolyte. Since we are primarily interested in a porous electrode theory model of a Li-ion cell in which the model of electrodes is one-dimensional we restrict our attention to such scenarios and leave a discussion of the general theory to \cite{richardson21}, where an homogenisation (\ie averaging) approach is adopted.}

\paragraph{Preliminaries.} The true electric potential is defined in terms of the electric field by the relation
\be 
\vect{E}=-\nabla \phi.
\label{dil-potential}
\ee
However the DFN model is formulated in the terms of $\vph$, the electric potential measured with respect to a lithium electrode, which is related to the true potential (see \cite{battrev} for details) via 
\be
\phi=\vph- \frac{RT}{F} \log \left(a_p \right) - \frac{\mubb_p^0 }{F}. 
\label{lipot}
\ee
where $a_p$ is the activity of the (positive) lithium ion.

\paragraph{Electrochemical and chemical potentials.}
The electrochemical potentials of the negative and positive ion species are, respectively given by 
\be
 \mub_{n}= \mubb_n- F\phi, \qquad
 \mub_p= \mubb_p+ F\phi. 
\label{echempot}
\ee
where $\mubb_n$ and $\mubb_p$ are the chemical potentials of the negative and positive ions respectively, and are usually written in terms of their activities, $a_n$ and $a_p$ respectively, as
\be
\mubb_n=\mubb_n^0 + RT \log(a_n), \qquad \mubb_p=\mubb_p^0 + RT \log(a_p).\label{chempot1}
\ee
It is also useful to define these two quantities in terms of $\G_e$, the Gibbs free energy of the electrolyte per unit volume,
\be
\mubb_n=\frac{\dd \G_e}{\dd n}, \qquad \mubb_p=\frac{\dd \G_e}{\dd p},  \label{chempot2}
\ee
where $n$ and $p$ are the concentrations of the negative and positive ions, respectively. It is common to refer to $\mubb_e$  the chemical potential of the electrolyte; this is defined by the relation
\be
\mubb_e=\frac{1}{2} (\mub_n+\mub_p)=\frac{1}{2} (\mubb_n+\mubb_p)
\ee
Notice that we can express $\mub_n$ and $\mub_p$ in terms solely of the measurable quantities $\mubb_e$ and $\vph$ as follows:
\be
\mub_p=F \vph, \qquad \mub_n= 2 \mubb_e-F  \vph.   \label{usefulreln}
\ee
Furthermore this is consistent with the definition of the chemical potential of the Li$^+$ in the active material (\ie $\mu_s=-F U_{eq,s}$) since the equilibrium condition for lithium ions, on either side of the particle interface, is equality of the electrochemical potentials for Li$^+$ in active material and electrolyte, \ie $-F U_{eq,s} +F \Phi_s= F \vph$, which is equivalent to the overpotential being zero, \ie $\eta=0$, as is to be expected when the intercalation reaction is in equilibrium.

%

\subsubsection{Derivation of the electrolyte energy conservation law}
We start by recalling that the averaged electrolyte equations, formulated in terms of conservation of Li$^+$ ions, are stated in \eqref{dfn1}. An alternative formulation of the first two of these equations, is provided in \eqref{dfn12}, which has been formulated in terms of the conservation of the negative counterion. In general the chemical energy density of the electrolyte is a function both  $p$, the concentration of Li$^+$ ions, and $n$, the concentration of the negative counterions, such that the Gibbs free energy of the electrolyte $\G_e(n,p)$ is a function of both these quantities;  and the chemical potentials of the ion species are thus $\mu_p=\dd \G_e/\dd p$ and $\mu_n=\dd \G_e/\dd n$. However, as is usual in an electrolyte, there is almost exact charge neutrality so that $n=p=c$, where $c$ is termed the electrolyte concentration. In such a charge neutral electrolyte filling a porous electrode, at volume fraction $\ev$, the chemical energy density (per unit volume) is $\ev \G_e(n,p)|_{n=p=c}$. The rate of change of this chemical energy density is thus
\be
\ev \frac{\dd \G_e}{\dd t}=\left. \ev \left(\frac{\dd \G_e}{\dd n}+\frac{\dd \G_e}{\dd p} \right) \right|_{{n=p=c}} \frac{\dd c}{\dd t}= \ev (\mubb_n +\mubb_p )\frac{\dd c}{\dd t}. \label{joe}
\ee
On noting that the definition of the electrochemical potentials \eqref{echempot} means that $(\mubb_n +\mubb_p )=(\mub_n +\mub_p )=2 \mu_e(c)$ it follow that we can rewrite \eqref{joe} in the form
\bes
\ev \frac{\dd \G_e}{\dd t}= \mub_n \left( \ev \frac{\dd c}{\dd t} \right)+ \mub_p \left( \ev \frac{\dd c}{\dd t} \right).
\ees
If we now substitute for the terms in the brackets from (\ref{dfn1}a) and (\ref{dfn12}a) we can rewrite the above in the form
\bes
\ev \frac{\dd \G_e}{\dd t}+ \mub_n \frac{\dd \avvqn}{\dd x} + \mub_p \frac{\dd \avvqp}{\dd x} =\mub_p \frac{\bet \avjn}{F}.
\ees
which in turn can be written in the form of the energy conservation equation
\be
\ev \frac{\dd \G_e}{\dd t}+ \frac{\dd}{\dd x} \left( \mub_n \avvqn+ \mub_p \avvqp \right)= \avvqn \frac{\dd \mub_n}{\dd x}+ \avvqp\frac{\dd \mub_p}{\dd x} +\mub_p \frac{\bet \avjn}{F}. \label{pluto}
\ee
In this equation we can identify $\mub_n \avvqn+ \mub_p \avvqp$ as the averaged flux of chemical energy, the first two terms on the right hand side as minus the averaged rate of heat production (per unit volume) in the electrolyte and the final term on the right hand side as the averaged rate of chemical energy flowing into the electrolyte per unit volume. This motivates us to rewrite \eqref{pluto} in the form
\be
\ev \frac{\dd \G_e}{\dd t}+ \frac{\dd}{\dd x}\avvqe = -\avvome +{\bet \avjn \vph}. \label{jupiter}
\ee
Here we have made of use of the fact that $\mub_p=F \vph$ and identified the averaged flux of chemical energy $\avvqe$ and the averaged rate of heat production per unit volume within the electrolyte $\avvome$ as follows:
\be
\avvqe =\mub_n \avvqn+ \mub_p \avvqp, \qquad \avvome=-\left(\avvqn \frac{\dd \mub_n}{\dd x}+ \avvqp\frac{\dd \mub_p}{\dd x} \right). \label{uranus}
\ee
In order to derive a more useful expression for the rate of heat production $\avvome$ we substitute for the averaged fluxes in (\ref{uranus}b) from (\ref{dfn1}b) and (\ref{dfn12}b), and after some rearrangement this results in the expression
\be
 \avvome= \De(c) \cd \frac{\dd c}{\dd x}  \frac{\dd }{\dd x} (\mub_n +\mub_p) +\frac{\avvj}{F} \left( (1-\tn)  \frac{\dd \mub_n}{\dd x}  -\tn\frac{\dd \mub_p}{\dd x}  \right). \label{neptune}
 \ee
We  note, that from the definition of $\avvj$ in (\ref{dfn1}d) and the relationships in \eqref{usefulreln}, that we can write
\be
\left((1-\tn)  \frac{\dd \mub_n}{\dd x} - \tn \frac{\dd \mub_p}{\dd x}   \right)= \frac{F}{\kappa(c) \cd} \avvj
\ee
and we can use this, together with the relation $\mub_n +\mub_p=2 \mu_e$, to rewrite the expression for the volumetric heat production in \eqref{neptune}, as follows:
\be
\avvome=2 \cd \De(c) \frac{d \mu_e}{d c} \left(\frac{\dd c}{\dd x} \right)^2+  \frac{1}{\kappa(c) \cd} \avvj^2. \label{saturn}
\ee

\paragraph{Summary of the energy conservation equations in electrolyte.} Collecting the various parts of the energy conservation equation in the electrolyte from  \eqref{jupiter}, \eqref{uranus} and \eqref{saturn} we arrive at the most concise formulation of the energy conservation equations, which reads
\be
 \frac{\dd}{\dd t} (\ev \G_e)+ \frac{\dd}{\dd x}\avvqe = -\avvome +{\bet \avjn \vph}, \label{mars1}\\
\avvqe =\mub_n \avvqn+ \mub_p \avvqp,\\
\avvome=2 \cd \De(c) \frac{d \mu_e}{d c} \left(\frac{\dd c}{\dd x} \right)^2+  \frac{1}{\kappa(c) \cd} \avvj^2.
\ee
Here $\ev \G_e$ is the chemical energy density in the electrolyte filling the porous electrode, $\avvqe$ is the flux of chemical energy, $\avvome$ is the averaged rate of loss of chemical energy (per unit volume) to heat and $\bet \avjn \vph$ is averaged rate of chemical energy (per unit volume) flowing into the electrolyte from the electrode particles.

\subsection{Energy conservation in the solid parts of the electrode matrices \label{solid}}

Electrical conduction within the solid parts of the anode and cathode are described by equations \eqref{dfn2}-\eqref{dfn3} and equations \eqref{dfn4}-\eqref{dfn5}, respectively.

A conservation equation for electrical energy within the solid part of the anode can be derived as follows.
Multiplying (\ref{dfn2}a) by $\Phi_a$ and integrating between $x=L_1$ and $x=L_2$ leads to the relation
\bes
-\int_{L_1}^{L_2} \bet(x) \avjn \Phi_a dx= \left[ j_a \Phi_a \right]_{L_1}^{L_2} -\int_{L_1}^{L_2}j_a\frac{\dd\Phi_a}{\dd x}  dx.
\ees
Then, on applying the boundary conditions \eqref{dfn3} to the first term on the right-hand side and substituting for $j_a$, from (\ref{dfn2}b), in the second term on the right-hand side we obtain the following energy balance
\be
-\int_{L_1}^{L_2} \bet(x) \avjn \Phi_a dx= -\frac{I \Phi_a|_{x=L_1}}{A}+ \int_{L_1}^{L_2} \sigma_a \left(\frac{\dd\Phi_a}{\dd x} \right)^2 dx. \label{bna5}
\ee
The term on the left-hand side of \eqref{bna5} is the rate of energy flow (per unit area) into the anode particles from the electrolyte, while the first term on the right-hand side is the rate of energy flow (per unit area) out of the solid part of the anode into the left-hand current collector and the final term is the rate of heating of the solid part of the anode (per unit area). In this way \eqref{bna5}  can be seen to be an energy conservation equation. An analogous relation can be derived, from \eqref{dfn4}- \eqref{dfn5}, for the solid part of the cathode; it is 
\be
-\int_{L_3}^{L_4} \bet(x) \avjn \Phi_c dx= \frac{I \Phi_c|_{x=L_4}}{A}+ \int_{L_3}^{L_4} \sigma_c \left(\frac{\dd\Phi_c}{\dd x} \right)^2 dx. \label{bna6}
\ee


\subsection{Energy Conservation equation in a single cell \label{whole}}
Here we consider a cell geometry described in \eqref{cell-geom} and illustrated in figure \ref{schematic}. In order to derive an energy conservation equation for the whole cell we first need to relate the particle number density $\N(x)$ to the BET surface area $\bet(x)$ and the particle radii $R_s(x)$. In order to do this we note that
\bes
\bet(x)=\left\{ \begin{array}{ccc} \ds \frac{3 \eps_{\rm part}(x)}{R_a(x)} & \mbox{in} & L_1<x<L_2 \\*[4mm]
\ds \frac{3 \eps_{\rm part}(x)}{R_c(x)} & \mbox{in} & L_3<x<L_4
\end{array}, \right.
\ees
where $\eps_{\rm part}$ is the volume fraction of electrode particles and that the particle number density  $\N(x)$ is given by 
\bes
\N(x)=\left\{ \begin{array}{ccc} \ds \frac{3 \eps_{\rm part}(x)}{4 \pi R_a^3(x)} & \mbox{in} & L_1<x<L_2 \\*[4mm]
\ds \frac{3 \eps_{\rm part}(x)}{4 \pi R_c^3(x)} & \mbox{in} & L_3<x<L_4
\end{array}. \right.
\ees
It follows that the particle number density  is related to the BET surface area by
\be
\N(x)=\left\{ \begin{array}{ccc} \ds \frac{\bet(x)}{4 \pi R_a^2(x)} & \mbox{in} & L_1<x<L_2 \\*[4mm]
\ds \frac{\bet(x)}{4 \pi R_c^2(x)} & \mbox{in} & L_3<x<L_4
\end{array}. \right. \label{Nb}
\ee

\paragraph{Chemical energy conservation in the active material}
We now seek an expression for the rate of change of the total amount of chemical energy stored within the active materials in both anode and cathode; that is we seek to determine
\bes
\frac{d}{d t} \left( \int_{L_1}^{L_2} \N(x) \ecsa(x,t) dx +\int_{L_3}^{L_4} \N(x) \ecsc(x,t) dx \right)
\ees
Before proceeding with this calculation it is helpful to rewrite the particle energy conservation equation \eqref{encons1}. By using the relation between the open circuit voltage $U_{eq}(c_s|_{r=R})$ and the overpotential $\eta$, namely $\eta=\Phi_s-\vph-U_{eq}(c_s|_{r=R})$, we can rewrite \eqref{encons1} in the form
\be
-\frac{d \ecs}{d t}= 4\pi R^2 \avjn \vph - 4 \pi R^2 \avjn \Phi_s+ 4 \pi R^2 \eta \avjn+ \omega^{\rm (heat)}_{\rm part,s}(x,t), \label{encons3}
\ee
Notably in this form of the energy conservation equation we can identify the first, second and third terms on the RHS of this equation as (i) the energy flux from the particle into the electrolyte, (ii) the energy flux from the electrode into the particle and (iii) the rate of heat production from the current flowing across the potential drop across the surface of the particle, respectively.

Having reformulated the energy conservation within the electrode particles in the form \eqref{encons3} the equation for the rate of change of the total amount of chemical energy stored within the active materials becomes
\be
-\frac{d}{d t} \left( \int_{L_1}^{L_2} \N(x) \ecsa(x,t) dx +\int_{L_3}^{L_4} \N(x) \ecsc(x,t) dx \right)=\int_{L_1}^{L_2} \N(x) \omega^{\rm (heat)}_{\rm part,a}(x,t) dx \non \\
+\int_{L_3}^{L_4}\N(x) \omega^{\rm (heat)}_{\rm part,c}(x,t) dx +  \int_{L_1}^{L_2} \bet(x)  \left(\avjn \vph -  \avjn \Phi_a+  \eta_a \avjn \right) dx \non \\
+\int_{L_3}^{L_4} \bet(x)  \left(\avjn \vph -  \avjn \Phi_a+  \eta_c \avjn \right) dx, \label{venus}
\ee
in which we have made use of the relation between $N(x)$ and $\bet(x)$ given in \eqref{Nb}.

We note: (I) the following expressions have been derived for the potential in the solid parts of the anode and cathode (in \eqref{bna5}-\eqref{bna6})
\be
\int_{L_1}^{L_2} \bet(x) \avjn \Phi_a dx= \frac{I}{A} \Phi_a|_{x=L_1}-\int_{L_1}^{L_2} \sigma_a \left( \frac{\dd \Phi_a}{\dd x} \right)^2 dx, \label{earth1} \\
 \int_{L_3}^{L_4} \bet(x) \avjn \Phi_c dx= -\frac{I}{A} \Phi_c|_{x=L_4}-\int_{L_3}^{L_4} \sigma_c \left( \frac{\dd \Phi_c}{\dd x} \right)^2 dx;\label{earth2} 
\ee
(II) that, since $\avjn=0$ in the separator in $L_2<x<L_3$, the following identity holds
\bes
\int_{L_1}^{L_2} \bet(x)  \avjn \vph dx+\int_{L_3}^{L_4} \bet(x)  \avjn \vph dx=\int_{L_1}^{L_4} \bet(x)  \avjn \vph dx;
\ees
and (III) that substitution of the electrolyte energy conservation law \eqref{mars1} into the right hand side of this expression, upon recalling that $\avvqe|_{x=L_1}=\avvqe|_{x=L_4}=0$, gives rise to the relation
\be
\int_{L_1}^{L_2} \bet(x)  \avjn \vph dx+\int_{L_3}^{L_4} \bet(x)  \avjn \vph dx=\int_{L_1}^{L_4} \left(\avvome + \frac{\dd}{\dd t} (\ev \G_e) \right)dx. \label{earth3} 
\ee

\paragraph{The energy conservation law.}
Substitution of \eqref{earth1}-\eqref{earth2} and \eqref{earth3} into \eqref{venus}  leads, after some rearrangement, to the desired energy conservation law
\be
-\frac{d}{d t} \left( \int_{L_1}^{L_2} \N(x) \ecsa(x,t) dx +\int_{L_3}^{L_4} \N(x) \ecsc(x,t) dx + \int_{L_1}^{L_4}\ev \G_e dx\right)= ~~~~~~~~~\non \\
\frac{I}{A} ( \Phi_c|_{x=L_4} - \Phi_a|_{x=L_1})+\int_{L_1}^{L_4} \avvome dx\non~~~~~~~~~~~~~~~~~~~~~~~~~~~~~~~~~~~~~~~~~~ \\
+ \int_{L_1}^{L_2} \N(x) \omega^{\rm (heat)}_{\rm part,a}(x,t) dx +\int_{L_1}^{L_2}\bet(x)\eta_a \avjn  dx+ \int_{L_1}^{L_2} \sigma_a \left( \frac{\dd \Phi_a}{\dd x} \right)^2 dx \non \\
 + \int_{L_3}^{L_4} \N(x) \omega^{\rm (heat)}_{\rm part,c}(x,t) dx +\int_{L_3}^{L_4}\bet(x)\eta_c \avjn dx  + \int_{L_3}^{L_4}\sigma_c \left( \frac{\dd \Phi_c}{\dd x} \right)^2 dx. \label{econs10}
 \ee
 where  
\be
\N(x)=\left\{ \begin{array}{ccc} \ds \frac{\bet(x)}{4 \pi R_a^2(x)} & \mbox{in} & L_1<x<L_2 \\*[4mm]
\ds \frac{\bet(x)}{4 \pi R_c^2(x)} & \mbox{in} & L_3<x<L_4
\end{array}. \right.
\ee
and
\be
\ecsa(x,t)&=&-\int_0^{R_a(x)} 4 \pi r^2 \left( \int F U_{eq,a}(c_a) d c_a \right) dr, \\
\ecsc(x,t)&=&-\int_0^{R_c(x)} 4 \pi r^2 \left( \int F U_{eq,c}(c_c) d c_c \right) dr,\\
\avvome&=&2 \cd \De(c) \frac{d \mu_e}{d c} \left(\frac{\dd c}{\dd x} \right)^2+  \frac{1}{\kappa(c) \cd} \avvj^2,\\
\G_e(c) &=& \int 2 \mu_e(c) dc. \label{econs15}
\ee
The energy conservation equations \eqref{econs10}-\eqref{econs15} that have been derived in this section can be seen to equivalent to the statement of the energy conservation law given in Results section (\S \ref{encons}) in equations \eqref{econs-fin}-\eqref{econs5}.

\section{Conclusions} 

\ma{In this work we have formally derived and validated, an energy conservation law  \eqref{echem}-\eqref{econs4}, for lithium ion batteries, from the Doyle-Fuller-Newman (DFN) model \eqref{dfn1}-\eqref{dfn11}. The significance of this result is twofold: (i) it highlights the fact that most, if not all, other works that purport to calculate heating, associated with irreversible energy losses, from the DFN model neglect important sources of energy dissipation within the cell and (ii) computations of energy dissipation within a cell provide a sound basis on which to optimise cell design, particularly as the formulation of the energy conservation law allows energy losses to be located to particular components of the cell. We were also able to derive, in \eqref{rev_heat}, an exact expression for the reversible heating in the cell that has not  previously appeared in the literature. However, we noted that for moderate discharges it should be well-approximated by the standard formula appearing in the literature. The rigorous mathematical approach that we have adopted in conducting this analysis has distinct advantages over more commonly adopted thermodynamic approaches, as it applies specifically to the DFN model, and therefore removes any doubt 
about the validity of the results when applied to that model. Furthermore, since a properly parametrised DFN model yields remarkably good voltage-current predictions, the irreversible energy losses that are computed from it, by using this rigorous approach, should  show the same level of fidelity to experiment.}

%

\paragraph{Acknowledgements}  GR and IK were supported by the Faraday Institution Multi-Scale Modelling (MSM) project Grant number EP/S003053/1.

\bibliographystyle{abbrv}
\bibliography{batterybiblio}

\end{document}